\documentclass[12pt,a4paper]{article}

\newcommand\np{\mbox{NP}}
\newcommand\p{\mbox{P}}
\newtheorem{theorem}{Theorem}[section]

\newtheorem{corollary}{Corollary}[section]

\title{About the finding of independent vertices of a graph}
\author{Anatoly D. Plotnikov}
\date{}

\begin{document}
\maketitle

\begin{abstract}
We examine the Maximum Independent Set Problem in an undirected graph.

The main result is that this problem can be considered as the solving
the same problem in a subclass of the weighted normal twin-orthogonal graphs. 

The problem is formulated which is dual to the problem above. It is 
shown that, for trivial twin-orthogonal graphs, any of its maximal
independent set is also maximum one.
\end{abstract}

\section{Statement of the problem} 

Consider the class $L$ of undirected graphs without loops and
multiple edges with weighted vertices.

Assume that there is a graph $G=(X, \Gamma, M)\in L$, where 
$X=\{x_{1},\ldots, x_{n}\}$ 
be the set of the graph vertices, $\Gamma$ is the mapping $X$ into $X$, and 
$M=\{\mu(x_{1}),\ldots, \mu(x_{n})\}$ is the set of the non-negative 
integers -- weights of the graph vertices. If 
$X_{1}=\{x_{i_{1}},\ldots, x_{i_{m}}\}\subset X$ then
$\Gamma X_{1}=\Gamma x_{i_{1}} \cup \cdots \cup \Gamma x_{i_{m}}$.

A graph $G= (X, \Gamma, M)\in L$ is called {\it isometric} if $\mu(x_{i}) 
=\mu(x_{j})$ ($i\not =j$) for all $x_{i}, x_{j}\in X$.

For any $A\subset X$ we shall designate
\[
\mu(A)= \sum_{\forall x_{i}\in A} \mu(x_{i}).
\]

As a {\it problem $Z$}, given on a graph $G=(X, \Gamma, M)\in L$, we shall 
call the problem of finding of vertex set $U\subset X$ such that satisfies 
conditions
\begin{equation}
\label{1.1}
U\cap \Gamma U-\oslash,
\end{equation}
\begin{equation}
\label{1.2}
U\cup \Gamma U=X
\end{equation}
and supplies the maximum of a function
\begin{equation}
\label{1.3}
\mu(U).
\end{equation}

Any vertex set $U\subset X$, satisfying the condition (\ref{1.1}), is called
{\it independent}. An independent set $U\subset X$, satisfying (\ref{1.2}), 
is called the {\it maximal independent set (MIS)} of the graph $G$.

A MIS ${\hat U}\subset X$, supplying the maximum of the function (\ref{1.3}), 
is called the {\it maximum independent set (MMIS)} of the graph $G$ 
(the optimum solution of the problem $Z$).

The problem $Z$ has the different applications \cite{christofides,
noskov, starobinets, gorbatov}. It has the special significance in 
Computation Complexity Theory, as it is \np-complete \cite{gary-johnson}. 
From the point of view of applications, the significance of any \np-complete 
problem is that it can be considered as a mathematical model of all discrete 
problems.

The existing methods for solving of the problem $Z$ (as a rule, in an isometric 
graph) consist in basic in finding all MIS of the graph $G\in L$ and
selection of them the maximum independent set \cite{christofides, gorbatov}.

The inefficiency of such approach to solving the problem $Z$ is proved by 
that the maximum number of the MISs $\sigma_{G}(n)$, a graph $G\in L$ can has, 
is equal to $\sigma_{G}(n)=\gamma(s)\dot 3^{r-1}$ \cite{christofides, 
gorbatov}, where $n=3r+s$, $\gamma(0)=3$, $\gamma(1)=4$, $\gamma(2)=6$. Hence, 
complexity of any algorithm, based on searching of all MIS of a graph $G$, can 
not have an evaluation better than $O(3^{n/3})$.

With the problem $Z$, given on a graph $G$, it is usually connected a problem 
of finding the maximum complete subgraph (the maximal clique) in the additional 
graph ${\overline G}= (X, {\overline \Gamma}, M)\in L$ as the subset of 
vertices ${\hat U}\in X$, inducing the maximum clique of a graph 
${\overline G}$, is MMIS of a graph $G$ \cite{christofides}.

Notice that the Maximum Clique Problem is a maximize problem, and from the 
point of view of the approach, accepted in Operations Research, is not dual 
to the problem $Z$.

Unfortunately, the difficulties, connected with finding MMIS of a graph 
$G\in L$, can not are overcome by development of a polynomial algorithm, 
enabling to find approximate solution of the problem $Z$ with a guaranteed 
deviation from the optimum solution \cite{gary-johnson}. Therefore, for 
development of the solution methods of the problem $Z$, it is necessary 
either to try to create an algorithm, discovering its exact solution (in this 
case it will be proven that \p = \np), or to find the exact solution of the 
problem for separate subclasses of graphs of $L$ (the majority authors go 
to the last way).

The main result of the given work is that the problem $Z$, given on an
arbitrary graph $G\in L$, can be considered as the solving the same problem in 
subclass of the normal conjugate-orthogonal graphs. A problem is formulated 
that is dual to the problem $Z$. It is shown that, for trivial 
conjugate-orthogonal graphs, any of its MIS is also a MMIS.

\section{A normal graph}

Divide a set of all vertices of a graph $G = (X, \Gamma,Œ)\in L$ into classes 
$Š_{j}$ ($j = \overline {1,s}$) such that if $å_{i_{1}}$, $å_{i_{2}}\in K_{j}$, 
then $\Gamma å_{i_{1}} = \Gamma å_{i_{2}}$. The set of all such classes of the
graph is designated by $H_{G}$.

\begin{theorem}
\label{t2.1} 
If $å_{i_{1}}\in U$ and $å_{i_{1}}\in Š_{j}$ then $Š_{j}\in U$, where $U$ be 
a MIS of a graph $G=(X, \Gamma,Œ)$.
\end{theorem}

Assume the conditions of Theorem \ref{t2.1} are satisfied and we allow 
that there exists a vertex $å_{i_{2}}\in Š_{j}$ such that $å_{i_{2}}\not\in U$ 
($i_{1}\not= i_{2}$).

As $\Gamma x_{i_{2}} = \Gamma x_{i_{1}}$ then $å_{i_{2}}\not\in \Gamma U$ owing 
to (\ref{1.1}). But, it takes into considering (\ref{1.2}), we have 
$å_{i_{2}}\in U$. The contradiction have obtained.\hfill Q.E.D.

\begin{theorem}
\label{t2.2} 
For any vertex $å_{i}\in •$ ($i =\overline {1, n}$) of a graph $G=(X,\Gamma,Œ)$
\[
\Gamma x_{i}=\bigcup_{r} K_{j_{r}}\qquad  (K_{j_{r}}\in H_{G})
\]
\end{theorem}

It is clear that the vertex set $\Gamma x_{i}$ can be divided into the classes 
$K_{j_{1}}^{\prime}$, \ldots , $Š_{j_{t}}^{\prime}$ as it is mentioned above. 
Assume that these classes of vertices are distinct from similar vertex
classes of the graph $G$, that is, 
$Š_{j_{r}}^{\prime}\subset Š_{j_{r}}$ and $Š_{j_{r}}^{\prime}\not=
Š_{j_{r}}$ ($£ =\overline {1, t}$).

It follows from here that there are the vertices 
$å_{k_{1}}\in Š_{j_{r}}^{\prime}\subset K_{j_{r}}$ and 
$x_{k_{2}}\in Š_{j_{r}}\setminus Š_{j_{r}}^{\prime}$ such that 
$å_{k_{1}}\in \Gamma å_{i}\in \Gamma x_{i}$ and 
$å_{k_{2}}\not\in \Gamma å_{i}$.

As vertices $å_{k_{1}}$, $å_{k_{2}}\in Š_{j_{r}}$ then 
$\Gamma å_{k_{1}} = \Gamma å_{k_{2}}$ by the definition. If 
$å_{k_{1}}\in \Gamma x_{i}$ then $å_{i}\in \Gamma å_{k_{1}}$, it signifies, 
$å_{i}\in \Gamma å_{k_{2}}$. Then we have $x_{k_{2}}\in \Gamma x_{i}$. 
The contradiction have obtained.\hfill Q.E.D.

Thus, it is established that $\Gamma x_{i}$, for any vertex $å_{i}\in •$ 
($i =\overline {1, ¯}$) of a graph $G=(X, \Gamma,Œ)$, be an union of some 
classes $Š_{j_{r}}\in H_{G}$.

\begin{corollary}
For any class $Š_{j}\in H_{G}$ of a graph $G=(X, \Gamma,Œ)$
\[
\Gamma K_{j}=\bigcup_{r} K_{j_{r}}\qquad (K_{j_{r}}\in H_{G})
\]
\end{corollary}

A graph $G_{1}=(X_{1}, \Gamma_{1},Œ_{1})\in L$ is called {\it normal} if for 
any two vertices $y_{j_{1}}$, $ã_{j_{2}}\in X_{1}$ ($j_{1}\not= j_{2}$) 
the relation $\Gamma y_{j_{1}}\not= \Gamma ã_{j_{2}}$ takes place.

Obviously, that for any graph $G=(X, \Gamma, M)\in L$ can be found a
mapping $\phi$: $G_{1}=\phi(G)$, where $G_{1}=(X_{1}, \Gamma_{1}, M_{1})\in L$ 
be the normal graph. Thus, $G_{1}=\phi(G)$ if $y_{j}=\phi (K_{j})$
($K_{j}\in H_{G}$) and $\Gamma_{1} y_{j}=\phi (\Gamma K_{j})$, 
$\mu (y_{j}) =\mu (K_{j})$ for all $y_{j}\in X_{1}$ ($j=\overline {1, s}$).

\begin{figure}[htbp]
\begin{center}
\unitlength 1.00mm
\linethickness{0.4pt}
\begin{picture}(82.00,55.11)
\put(20.00,10.00){\line(-1,1){10.00}}
\put(30.00,20.00){\line(-1,-1){10.00}}
\put(10.00,20.00){\line(0,1){20.00}}
\put(10.00,40.00){\line(1,1){10.00}}
\put(20.00,50.00){\line(1,-1){10.00}}
\put(30.00,40.00){\line(0,-1){20.00}}
\put(30.00,20.00){\line(-2,1){20.00}}
\put(10.00,30.00){\line(1,0){20.00}}
\put(30.00,30.00){\line(-2,-1){20.00}}
\put(10.00,30.00){\line(2,1){20.00}}
\put(10.00,40.00){\line(2,-1){20.00}}
\put(20.00,10.00){\line(0,1){20.00}}
\put(20.00,50.00){\circle*{2.50}}
\put(10.00,40.00){\circle*{2.50}}
\put(30.00,40.00){\circle*{2.50}}
\put(10.00,30.00){\circle*{2.50}}
\put(20.00,30.00){\circle*{2.50}}
\put(30.00,30.00){\circle*{2.50}}
\put(10.00,20.00){\circle*{2.50}}
\put(30.00,20.00){\circle*{2.50}}
\put(20.00,10.00){\circle*{2.50}}
\put(54.00,30.00){\line(1,1){10.00}}
\put(64.00,40.00){\line(1,0){15.00}}
\put(54.00,30.00){\line(1,-1){10.00}}
\put(64.00,20.00){\line(1,0){15.00}}
\put(80.00,20.00){\circle*{2.50}}
\put(64.00,20.00){\circle*{2.50}}
\put(54.00,30.00){\circle*{2.50}}
\put(64.00,40.00){\circle*{2.50}}
\put(80.00,40.00){\circle*{2.50}}
\put(24.67,6.44){\makebox(0,0)[cc]{$x_{8}$}}
\put(5.56,16.22){\makebox(0,0)[cc]{$x_{1}$}}
\put(4.44,31.78){\makebox(0,0)[cc]{$x_{2}$}}
\put(4.44,42.89){\makebox(0,0)[cc]{$x_{3}$}}
\put(21.11,55.11){\makebox(0,0)[cc]{$x_{4}$}}
\put(33.56,43.33){\makebox(0,0)[cc]{$x_{5}$}}
\put(34.22,32.00){\makebox(0,0)[cc]{$x_{6}$}}
\put(33.56,16.89){\makebox(0,0)[cc]{$x_{7}$}}
\put(19.11,39.56){\makebox(0,0)[cc]{$x_{9}$}}
\put(79.77,24.67){\makebox(0,0)[cc]{(1)}}
\put(66.00,24.22){\makebox(0,0)[cc]{(3)}}
\put(50.89,33.56){\makebox(0,0)[cc]{(2)}}
\put(65.77,35.56){\makebox(0,0)[cc]{(2)}}
\put(79.77,36.22){\makebox(0,0)[cc]{(1)}}
\put(69.55,30.22){\makebox(0,0)[cc]{$y_{2}=\{x_{2},x_{5}\}$}}
\put(58.22,45.00){\makebox(0,0)[cc]{$y_{3}=\{x_{3},x_{6}\}$}}
\put(85.00,45.00){\makebox(0,0)[cc]{$y_{4}=\{x_{4}\}$}}
\put(57.77,14.50){\makebox(0,0)[cc]{$y_{1}=\{x_{1},x_{7},x_{9}\}$}}
\put(85.00,14.50){\makebox(0,0)[cc]{$y_{5}=\{x_{8}\}$}}
\put(20.00,2.00){\makebox(0,0)[cc]{(a)}}
\put(68.00,2.00){\makebox(0,0)[cc]{(b)}}
\end{picture}
\caption{Normalization of the graph}
\label{h1}
\end{center}
\end{figure}
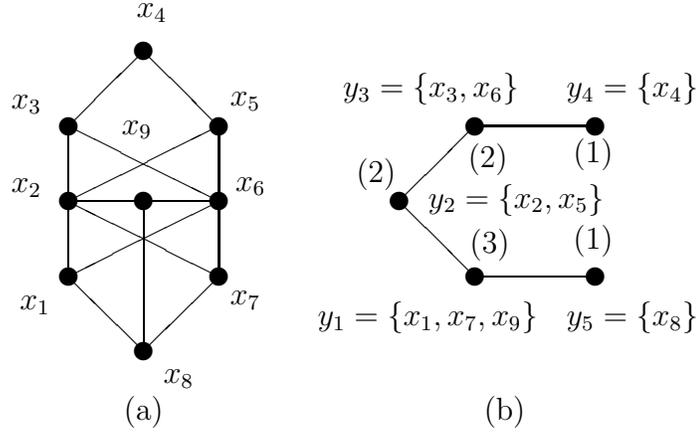

Fig. \ref{h1} (a) shows the graph $G\in L$ with the unit weights of its 
vertices, and Fig. \ref{h1} (B) shows the normal graph $G_{1}$ that corresponds
it (weights of its vertices are put in brackets).

Designate by $L_{H}$ the set of all normal graphs with the weighted vertices 
that correspond graphs of the class $L$.

Further, speaking about a graph $G=(X, \Gamma,Œ)$, we shall mean that 
$G\in L_{H}$. Besides, we assume that $Card X=n$.

\section{A twin-orthogonal graph}

Let $G= (X, \Gamma, Œ)\in L_{H}$.

The adjacent vertices $å_{1}$, $x_{2}\in •$ of the graph $G$ is called 
{\it orthogonal} if for all vertices 
$å_{i_{1}}\in \Gamma x_{1}\setminus \{x_{2}\}$ and 
$x_{i_{2}}\in \Gamma å_{2}\setminus \{x_{1}\}$, when they exist, the 
relation are fulfilled:

\begin{equation}
\label{3.1}
\Gamma x_{1}\subset \Gamma x_{i_{2}},\quad \Gamma x_{2}\subset \Gamma x_{i_{1}}.
\end{equation}

\begin{theorem}
\label{t3.1}
If at least one of adjacent vertices $å_{1}$, $x_{2}\in X$ of a graph $G$ is
dangling then the vertices $x_{1}$ and $x_{2}$ are orthogonal.
\end{theorem}

Really, suppose, for example, a vertex $x_{1}\in •$ of a graph $G$, adjacent 
with a vertex $å_{2}\in X$, is dangling. Hence, $\Gamma å_{1} = \{å_{2}\}$.

Then we shall have $\Gamma å_{1}\setminus \{x_{2}\}\not= \oslash$ and 
$\Gamma å_{1}\subset \Gamma x_{i_{2}}$ for all 
$x_{i_{2}}\in \Gamma å_{2}\setminus \{å_{1}\}$ when
$\Gamma x_{2}\setminus \{x_{1}\}\not= \oslash$ 
(as $x_{2}\in \Gamma x_{i_{2}}$).\hfill Q.E.D.

\begin{theorem}
\label{t3.2}
Let $U\subset •$ be an arbitrary MIS of a graph $G=(X, \Gamma,M)$. If 
$x_{1}$, $x_{2}\in •$ is the orthogonal vertices of $G$ either $å_{1}\in U$ 
or $å_{2}\in U$.
\end{theorem}

Assume that the conditions of Theorem \ref{t3.2} are satisfied, and 
suppose that $x_{1}$, $x_{2}\in \Gamma U$. Then there exists at least vertex 
$å_{3}\in \Gamma x_{1}$ such that $x_{3}\in U$, and at least vertex 
$å_{4}\in \Gamma x_{2}$ such that $x_{4}\in U$.

By the condition (\ref{3.1}), for the orthogonal vertices $x_{1}$, $x_{2}\in X$, 
we shall have $x_{3}\in \Gamma å_{4}$ and $x_{4}\in \Gamma x_{3}$, that is, 
the vertices $x_{3}$, $x_{4}\in U$ are adjacent. We have received the
contradiction.\hfill Q.E.D.

A graph ${\tilde G}=({\tilde X}, {\tilde \Gamma}, {\tilde M})\in L_{H}$ is
called {\it twin-orthogonal} if graph vertices can divide into pairs of the
orthogonal vertices. It is clear that $' rd (•)=n=2k$, where $k$ is a
non-negative integer.

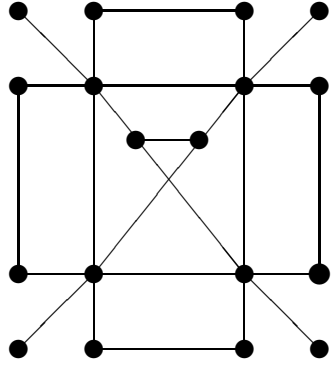
\begin{figure}[htbp]
\begin{center}
\unitlength 1.00mm
\linethickness{0.4pt}
\begin{picture}(51.35,56.31)
\put(20.00,20.00){\line(0,1){25.00}}
\put(20.00,45.00){\line(1,0){20.00}}
\put(40.00,45.00){\line(0,-1){25.00}}
\put(40.00,20.00){\line(-1,0){20.00}}
\put(20.00,20.00){\line(4,5){20.00}}
\put(40.00,45.00){\line(0,1){10.00}}
\put(40.00,55.00){\line(-1,0){20.00}}
\put(20.00,55.00){\line(0,-1){10.00}}
\put(20.00,45.00){\line(4,-5){20.00}}
\put(40.00,20.00){\line(1,0){10.00}}
\put(50.00,20.00){\line(0,1){25.00}}
\put(50.00,45.00){\line(-1,0){10.00}}
\put(40.00,45.00){\line(1,1){10.00}}
\put(20.00,45.00){\line(-1,0){10.00}}
\put(10.00,45.00){\line(0,-1){25.00}}
\put(10.00,20.00){\line(1,0){10.00}}
\put(20.00,20.00){\line(0,-1){10.00}}
\put(20.00,10.00){\line(1,0){20.00}}
\put(40.00,10.00){\line(0,1){10.00}}
\put(40.00,20.00){\line(1,-1){10.00}}
\put(20.00,20.00){\line(-1,-1){10.00}}
\put(20.00,45.00){\line(-1,1){10.00}}
\put(25.87,37.78){\line(1,0){8.33}}
\put(10.00,55.00){\circle*{2.50}}
\put(20.00,55.00){\circle*{2.50}}
\put(40.00,55.00){\circle*{2.50}}
\put(50.00,55.00){\circle*{2.50}}
\put(10.00,45.00){\circle*{2.50}}
\put(20.00,45.00){\circle*{2.50}}
\put(40.00,45.00){\circle*{2.50}}
\put(50.00,45.00){\circle*{2.50}}
\put(25.56,37.78){\circle*{2.50}}
\put(34.02,37.78){\circle*{2.50}}
\put(10.00,20.00){\circle*{2.50}}
\put(20.00,20.00){\circle*{2.50}}
\put(40.00,20.00){\circle*{2.50}}
\put(50.00,20.00){\circle*{2.70}}
\put(10.00,10.00){\circle*{2.50}}
\put(20.00,10.00){\circle*{2.50}}
\put(40.00,10.00){\circle*{2.50}}
\put(50.00,10.00){\circle*{2.50}}
\end{picture}
\caption{A twin-orthogonal graph}
\label{h2}
\end{center}
\end{figure}

Fig. \ref{h2} is represented of a twin-orthogonal graph with the unit weights 
of the vertices.

We shall be to say that the twin-orthogonal graph 
${\tilde G} = ({\tilde X},{\tilde \Gamma}, {\tilde Œ})$ {\it corresponds  
a graph} $G = (X, \Gamma,Œ)$, if:

\begin{itemize}
\item[a)] $X\subset {\tilde X}$;
\item[b)] $\mu (x_{i})=0$ for any vertex $x_{i}\in {\tilde X}\setminus X$;
\item[c)] any MIS $U\subset X$ of $G$ can be obtained from some MIS 
${\tilde U}\subset {\tilde X}$ of ${\tilde G}$ by removal of all vertices 
$x_{i}\in {\tilde X}$ such that $x_{i}\not\in X$.
\end{itemize}

It is easy to see that one of twin-orthogonal graphs 
${\tilde G} = ({\tilde X},{\tilde \Gamma}, {\tilde Œ})$, corresponding
a graph $G=(X, \Gamma,M)$, can be constructed as follows.

Let $X_{1}\subset •$ be the set of all vertices of the graph $G$, not being 
orthogonal for one vertex of this graph. We join a set of vertices 
$•_{2}=\{x_{n+1}, \ldots , x_{n+p}\}$ ($p= ' rd (X_{1})$) to the graph $G$, 
and each of vertices $å_{k}\in •_{2}$ we connect by an edge with one and only 
one of vertices $å_{j}\in X_{1}$. We assume that $\mu (x_{k})=0$ for all 
$å_{k}\in X_{2}$.

It is clear that, as a result, a twin-orthogonal graph 
${\tilde G} = ({\tilde X},{\tilde \Gamma}, {\tilde Œ})$ will be obtained that
is induced on a vertex set ${\tilde •}=•\cup •_{2}$, where 
${\tilde \Gamma} x_{i}= \Gamma å_{i}$ for any vertex 
$å_{i}\in •\setminus •_{1}$, 
${\tilde \Gamma x_{j}}= \Gamma x_{j}\cup \{å_{k}\}$ for all $x_{j}\in •_{1}$ 
and ${\tilde \Gamma} x_{h}=\{x_{j}\}$ for all $x_{k}\in X_{2}$.

It is easy to be convinced that the constructed twin-orthogonal graph 
${\tilde G}$ corresponds the initial graph $G$.

More simple way for a construction of the twin-orthogonal graph 
${\tilde G}$ = $({\tilde X}$, ${\tilde \Gamma}$, ${\tilde Œ})$, corresponding 
a graph $G$ = $(X, \Gamma,M)$, is based on Theorem \ref{t3.1}. 

We shall join a vertex set $•_{1}$ ($' rd (X_{1}) = ' rd (X)$) to a graph $G$ 
such that each vertex $å_{j}\in •_{1}$ we shall connect by an edge with one 
and only one of vertices $å_{i}\in X$. We assume
$\mu (å_{j})=0$ for all $å_{j}\in X_{1}$. As a result, obviously, it be also 
obtained a twin-orthogonal graph 
${\tilde G} = ({\tilde X},{\tilde \Gamma}, {\tilde Œ})$
induced on a set of vertices ${\tilde •}=• \cup X_{1}$, where 
${\tilde \Gamma} å_{i}= \Gamma å_{i}\cup \{å_{j}\}$ for any vertex 
$å_{i}\in •$ and ${\tilde \Gamma} x_{j}= \{å_{i}\}$ for all $å_{j}\in X_{1}$.

\begin{theorem}
\label{t3.3} 
If ${\tilde U}\subset {\tilde •}$ be an optimum solution of the problem $Z$ 
on the twin-orthogonal graph ${\tilde \Gamma} å_{i}= \Gamma å_{i}\cup \{å_{j}\}$
corresponding a graph $G=(X, \Gamma, Œ)$ then an optimum solution
${\hat U}\in •$ of the problem $Z$, given on the graph $G$, can be obtained by 
removal from ${\tilde U}$ of all vertices $å_{i}\in {\tilde U}$ such that 
$å_{i}\in {\tilde •}\setminus •$, and, besides, 
$\mu ({\hat U})=\mu ({\tilde U})$.
\end{theorem}

It follows from the definition of a twin-orthogonal graph ${\tilde G}$,
corresponding a graph $G$.\hfill Q.E.D.

\section{Some properties of a twin-orthogonal graph}

Let $L_{0}$ be the set of the normal twin-orthogonal graphs.

\begin{theorem}
\label{t4.1}
If $U_{1}$, $U_{2}\in •$ be the different MISs of a twin-orthogonal graph
$G=(X, \Gamma, Œ)\in L_{0}$ then $' rd (U_{1})=' rd (U_{2})=k$, where 
$' rd (•)=n=2k$, $k$ be a non-negative integer.
\end{theorem}

It follows from Theorem \ref{t3.2}.\hfill Q.E.D.

A twin-orthogonal graph $G=(X, \Gamma, Œ)$ is called {\it trivial} if for any 
orthogonal vertices $å_{i}$, $å_{j}\in X$ the relation is fulfilled: 
$\mu (å_{i})=\mu (x_{j})$.

\begin{theorem}
\label{t4.2}
If $G=(X, \Gamma, Œ)$ be a trivial twin-orthogonal graph then any MIS is also 
MMIS.
\end{theorem}

It follows from Theorems \ref{t3.2} and \ref{t4.1}.\hfill Q.E.D.

\begin{theorem}
\label{t4.3}
If $x_{1}$, $x_{2}\in •$ be the orthogonal vertices of a graph $G=(•,\Gamma,Œ)$
then any pair of vertices from a set 
$\Gamma x_{1}\setminus \{x_{2}\}\not= \oslash$  
($\Gamma x_{2}\setminus \{x_{1}\}\not= \oslash$) is not orthogonal.
\end{theorem}

Assume that the conditions of Theorem \ref{t4.3} are satisfied, and we 
suppose that the vertices $å_{i_{1}}$, 
$å_{i_{2}}\in \Gamma x_{1}\setminus \{å_{2}\}$ are orthogonal. Then we have 
$å_{1}\in \Gamma x_{i_{1}}$, and $å_{1}\in \Gamma å_{i_{2}}$, that is,
the relation (\ref{3.1}) are not fulfilled for vertices $x_{i_{1}}$, 
$x_{i_{2}}\in •$. We have obtained the contradiction.\hfill Q.E.D.

\begin{corollary}
If the vertices $å_{1}$, $x_{2}\in •$ of a graph $G=(X, \Gamma,Œ)$ are
orthogonal then they do not form a three-vertex clique with any vertex
$å_{i}\in •$ ($i\not= 1$, $i\not= 2$).
\end{corollary}

\begin{corollary}
If the vertices $å_{1}, å_{2}\in •$ of a graph $G=(X, \Gamma,Œ)$ are
orthogonal then 
$(\Gamma x_{1}\setminus \{x_{2}\})\cap 
(\Gamma x_{2}\setminus \{x_{1}\})=\oslash$.
\end{corollary}

\section{A dual problem}

Further, for convenience, any two orthogonal vertices of a graph
$G$ = $(X$, $\Gamma,Œ)$ $\in L_{0}$ we shall designate by $å_{i}$ and $å_{i}^{*}$.

A graph $G^{*}=(X, \Gamma^{*},M)\in L_{0}$, obtained from a graph 
$G=(X, \Gamma,Œ)\in L_{0}$ by renaming of pairs of orthogonal vertices, 
is called {\it conjugate} for the graph $G$.

Thus, any orthogonal vertices $å_{i}$, $å_{i}^{*}\in •$ are adjacent in graphs
$G$ and $G^{*}$. The vertices $å_{i}$, $å_{j}\in X$, if they are not orthogonal
in the graph $G$, are adjacent in the graph $G^{*}$ if and only if 
corresponding vertices $x_{i}^{*}$, $x_{j}^{*}\in X$ are adjacent in the 
graph $G$.

Obviously, that $(G^{*})^{*}=G$.

A problem of finding of a vertex set $U\subset •$ of a graph 
$G^{*}=(X, \Gamma^{*},Œ)$, satisfying conditions (\ref{1.1}), (\ref{1.2}) 
and supplying the minimum of the function (\ref{1.3}), we shall call {\it dual}
to the problem $Z$.

A MIS ${\check U}\subset X$, supplying the minimum of the function (\ref{1.3}), 
is called the {\it minimum independent set of vertices (MNMIS)} of a graph 
$G^{*}$.

The following statements are proved easily.

\begin{theorem}
\label{t5.1}
If $U\subset •$ be a MIS of a graph $G=(X, \Gamma,M)$ then 
$\Gamma U = •\setminus U$ be a MIS of the conjugate graph 
$G{*}=(X, \Gamma^{*},Œ)$.
\end{theorem}

\begin{theorem}
\label{t5.2}
If $U_{1}$, $U_{2}\in •$ be MISs of a graph $G=(X, \Gamma,Œ)$ then 
$\mu (U_{1})\geq \mu (U_{2})$ if and only if 
$\mu (\Gamma U_{1})\leq \mu (\Gamma U_{2})$.
\end{theorem}

The following theorem is a corollary of Theorems \ref{t5.1} and \ref{t5.2}.

\begin{theorem}
MIS ${\hat U}\subset •$ is MNMIS of a graph $G=(X, \Gamma,Œ)$ if and only if
$\Gamma {\hat U}=•\setminus {\hat U}$ is MNMIS of a conjugate graph
$G^{*}=(X, \Gamma^{*},Œ)$.
\end{theorem}

\begin{theorem}
Let ${\hat U}$, ${\check U}\subset •$ be MMIS and MNMIS of a graph 
$G$ = $(X$, $\Gamma$, $M)$ respectively. Then a relation takes place
\[
0\leq \mu ({\hat U})-\mu ({\check U})\leq 
\sum_{\forall x_{i}, x_{i}^{*}\in X} |\mu (x_{i})-\mu(x_{i}^{*})|.
\]
\end{theorem}

It is also easy to be convinced in a validity of this statement.


\begin{thebibliography}{1}

\bibitem{christofides}
N.~Christofides.
\newblock {\em Graph theory (An Algorithmic Approach)}.
\newblock Academic press, New York, 1975.

\bibitem{gary-johnson}
M.~R. Garey and D.~S. Johnson.
\newblock {\em Computers and Intractability}.
\newblock W.H.Freeman and Company, San Francisco, 1979.

\bibitem{gorbatov}
V.~A. Gorbatov.
\newblock {\em Foundation of Discrete mathematics (in Russian)}.
\newblock High school, Moscow, 1986.

\bibitem{noskov}
V.~V. Noskov.
\newblock The maximum independent sets and constructive coding problems.
\newblock {\em Problemy kibernetiki (in Russian)}, 36:33--54, 1979.

\bibitem{starobinets}
S.~M. Starobinets.
\newblock On an algorithm for finding the maximum stable set of a graph.
\newblock {\em Izvestia Akademii Nauk SSSR. Tehnicheskaya kibernetika (in
  Russian)}, (5):135--140, 1972.

\end{thebibliography}
\end{document}